\documentclass[final,twocolumn,5p]{elsarticle}

\usepackage[english]{babel}
\usepackage[babel]{csquotes}
\usepackage{natbib}

\usepackage{color}

\usepackage{amsmath}
\usepackage{amsfonts}
\usepackage{amssymb}

\def\be{\begin{equation}}
\def\ee{\end{equation}}

\usepackage{hyperref}

\journal{Physics Letters B}

\begin{document}

\begin{frontmatter}

\title{A conservative assessment of the current constraints on dark matter annihilation 
from Cosmic Rays and CMB observations}

\author[DIFA,INFN]{Nicol\`o Masi\corref{cor1}}
\ead{masi@bo.infn.it}

\author[DIFA,INAF,INFN]{Mario Ballardini}
\ead{ballardini@iasfbo.inaf.it}

\cortext[cor1]{Corresponding author}
\address[DIFA]{DIFA, Dipartimento di Fisica e Astronomia,
Alma Mater Studiorum Universit\`a di Bologna,\\
viale Berti Pichat 6/2, I-40127 Bologna, Italy}
\address[INFN]{INFN, Sezione di Bologna, \\
via Irnerio 46, I-40126 Bologna, Italy}
\address[INAF]{INAF/IASF-BO, Istituto di Astrofisica Spaziale e Fisica Cosmica di Bologna, \\
via Gobetti 101, I-40129 Bologna - Italy}

\begin{abstract}
In view of the current interest in combining different observations to constraint 
annihilating WIMP dark matter, we examine the relation between the Sommerfeld effect at the 
recombination epoch and in the galactic halo. By considering an up-to-date collection of 
interpolations of cosmic rays lepton data (AMS-02 2014, Fermi and PAMELA), as dark matter 
annihilation signals, we show that current cosmic rays measurements and recent 
Planck 2015 constraints from CMB anisotropies almost overlap for dark matter masses of the order 
of few $TeV$, although great theoretical uncertainties afflict cosmic rays and dark matter descriptions.
Combining cosmic rays fits we obtain proper minimal regions allowed by CMB observations, especially 
for $\mathit\mu$ and $\mathit\tau$ annihilation channels, once assumed viable values of the efficiency 
factor for energy absorption at recombination: the results are consistent with those obtained by the 
Planck collaboration but allow a slightly larger overlap between Cosmic Rays constraints 
from the lepton sector and CMB. Incoming AMS-02 measurements of cosmic rays antiprotons will help to 
clarify the conundrum.
\end{abstract}


\end{frontmatter}

\section{Introduction}
\label{sec:introduction}

AMS-02 2014 measurements of cosmic rays (CR) leptons 
\cite{Accardo:2014lma, Aguilar:2014fea, Aguilar:2014mma}, 
which have confirmed the rise of the positron fraction for kinetic energy above 
$\sim10\ GeV$, up to $\sim1\ TeV$, 
have stimulated different interpretations of this 
excess of positrons as primary evidence of dark matter (DM) annihilation. 
The interpretation of DM as a new source of positrons, to explain 
the departure from the pure secondary positron fraction, can be tested with other complementary and independent 
indirect measurements, as the cosmic microwave background (CMB) anisotropies.
In fact, if DM particles self-annihilate at a sufficient rate, the expected signal would be directly 
sensitive to their thermally averaged cross section and it could have drawn an imprint on CMB temperature and polarization anisotropies.

\begin{figure}[ht!!!]
\centering
\includegraphics[height=4cm]{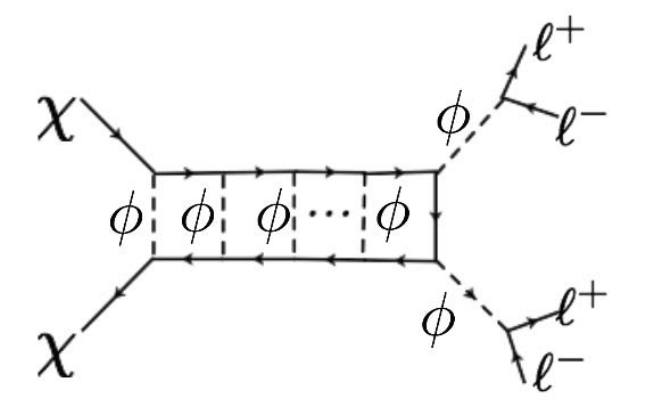}
\caption{Feynman diagram for the Sommerfeld enhancement induced by a scalar.}
\label{fig:Feynman}
\end{figure}

The aim of the present letter is to discuss the link 
between the DM annihilation cross sections at freeze-out, recombination and in the 
Milky Way galactic halo, which is essential to compare different indirect constraints on DM itself. 
The three different physical quantities are defined by the environment 
in three different epochs. 
The relation between them is not trivial, not a simply decreasing function 
and so it must be carefully analyzed to put coherent constraints on 
DM properties. Once defined a consistent framework, one can compare information from CR physics with CMB 
observations.\\
The need of a high thermally averaged annihilation cross section, not purely thermal, \textit{i.e.} 
$\langle\sigma\nu\rangle^\textup{ann} \gg 3\times 10^{-26}\ cm^3\ s^{-1}$, comes from the above mentioned observation of a huge 
excess in the CR positron fraction, in contrast with the expected behavior of secondaries 
produced in the interstellar medium (ISM). In order to describe this experimental 
evidence, a very high annihilation cross section has to be invoked, compared with typical 
expectation for a s-wave annihilating thermal relic matching the observed DM abundance 
\cite{Masi:2015fka}.\\
The main way to interpret 
the positron excess is the Sommerfeld enhancement \cite{Masi:2013phd}, 
a non-perturbative quantum effect which modifies the annihilation cross section in the regime 
of small relative velocity of the annihilating particles and in presence of an effectively 
long-range force between them. Indeed, this well-known quantum mechanical effect can occur in 
DM annihilations in the galactic halo, if the two annihilating particles exchange an 
interaction mediated by a force carrier. From a Feynman diagram point of view, the 
Weakly Interacting Massive Particles (WIMP) interact with the new boson through a multi-box 
diagram, depicted in Fig.~\ref{fig:Feynman}, annihilate into $\chi\chi\to\phi\phi$, and 
the decay of $\phi$ produces more light leptons than expected.

The structure of the paper is the following: in Section~\ref{sec:two} the basic setup 
of the Sommerfeld enhancement and the relation between the galactic halo cross section 
and the one at recombination are briefly reviewed; in Section~\ref{sec:three} we report 
a collection of CR best fits 
for TeV-ish DM candidates capable to reproduce the observed positron excess 
and we discuss the main sources of uncertainties (from CR physics and the DM sector), 
in order to compare predictions with 
CMB constraints and obtain some general remarks in Section~\ref{sec:conclusion}.

\section{The quantum Sommerfeld enhancement}
\label{sec:two}

The Sommerfeld enhancement is 
fundamental 
to interpret indirect DM searches.
The thermally averaged DM annihilation cross section (at any time) can be generally 
decomponed into powers of the velocity $\nu$ \cite{Masi:2013phd}:
\begin{align}
\label{eqn:sigma}
\langle\sigma\nu\rangle^\textup{ann}&=\sum_{n=0}^\infty c_n\langle\nu^{2n}\rangle+\text{Sommerfeld effect} \notag\\
&\sim A+B\langle\nu^2\rangle+C(\overline\nu+\langle\nu\rangle)^{-1}+\mathcal{O}(\langle\nu^4\rangle)
\end{align}
where $\nu$ is the relative velocity of the annihilating particles, that is 
$\nu=\nu_\textup{rel}=2\nu_\textup{CM}$, $A$ is the constant s-wave term, $B\nu^2$ is the p-wave 
term and $C(\overline\nu+\langle\nu\rangle)^{-1}$ the Sommerfeld term where $\overline\nu$ stands 
for a proper asymptotic (cut-off) value for the low velocity regime. The quadratic and quartic terms in 
Eq.~\eqref{eqn:sigma} are commonly neglected at a freeze-out description, whereas the so called 
Sommerfeld term is suppressed in the relativistic velocities regime \cite{Cirelli:2008pk}.\\
The enhancement of the s-wave for these velocities is defined by the ratio 
of the masses of the DM candidate and the Sommerfeld boson $\phi$ which provides the 
boosted annihilation, and by the effective coupling $g_\chi\approx\sqrt{4\pi\alpha_\chi}$. 
Defining the dimensionless parameter $\varepsilon_\phi^{-1}=\alpha_\chi m_\textup{DM}/m_\phi$, 
the condition for the enhancement is $\varepsilon_\phi^{-1}>1$, that is 
$m_\phi/m_\textup{DM} \lesssim 10^{-(3\div2)}$ and $\alpha_\chi \gtrsim 10^{-(2\div1)}$.
\begin{figure}[h!!!]
\centering
\includegraphics[height=6cm]{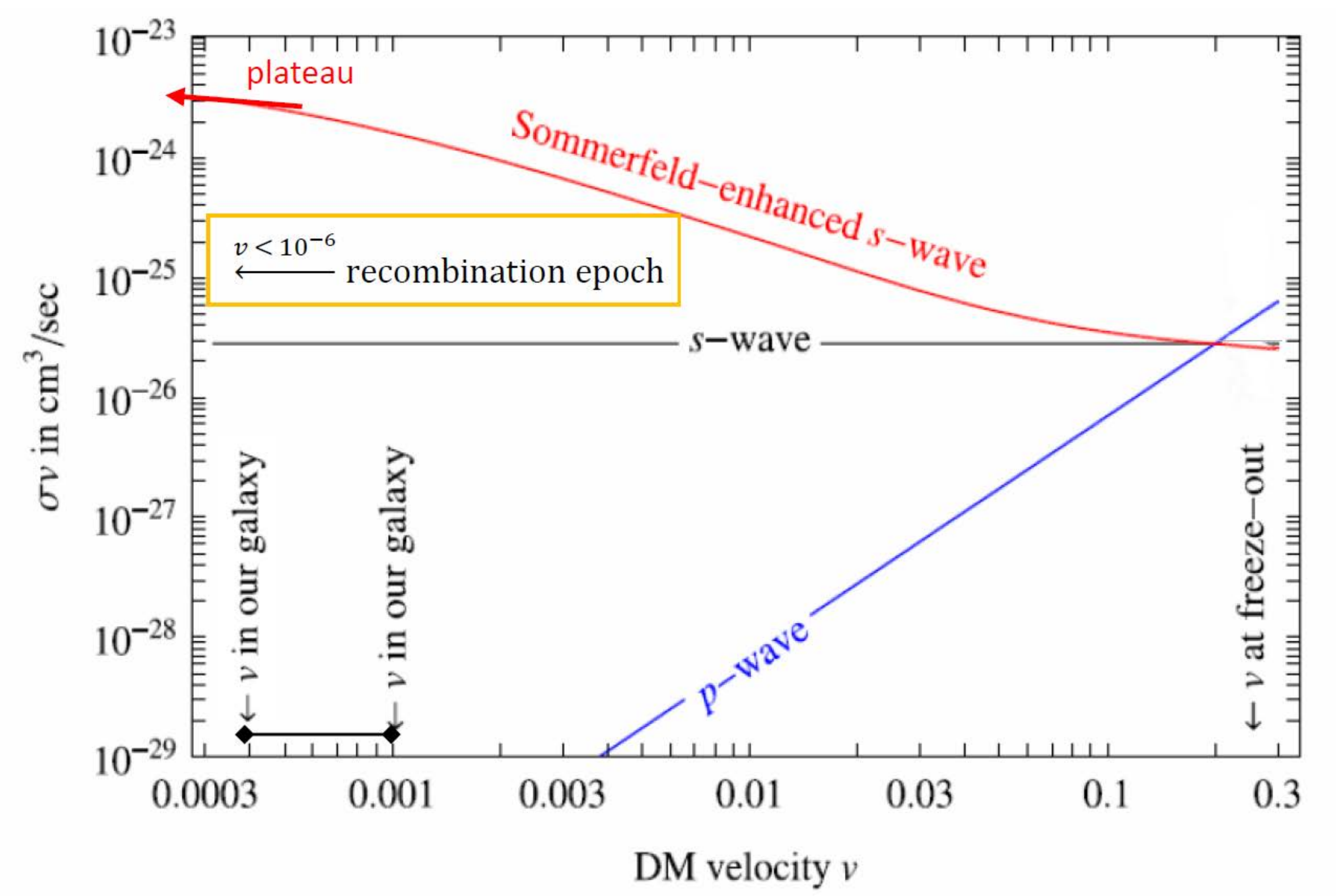}
\caption{The three main terms of the DM annihilation cross section as functions of $\beta$ 
($\nu$ in natural unit on the abscissa). See \cite{Cirelli:2008pk} for insights.}
\label{fig:CrossSection}
\end{figure}

\begin{figure}[htp]
\centering
\includegraphics[height=6cm]{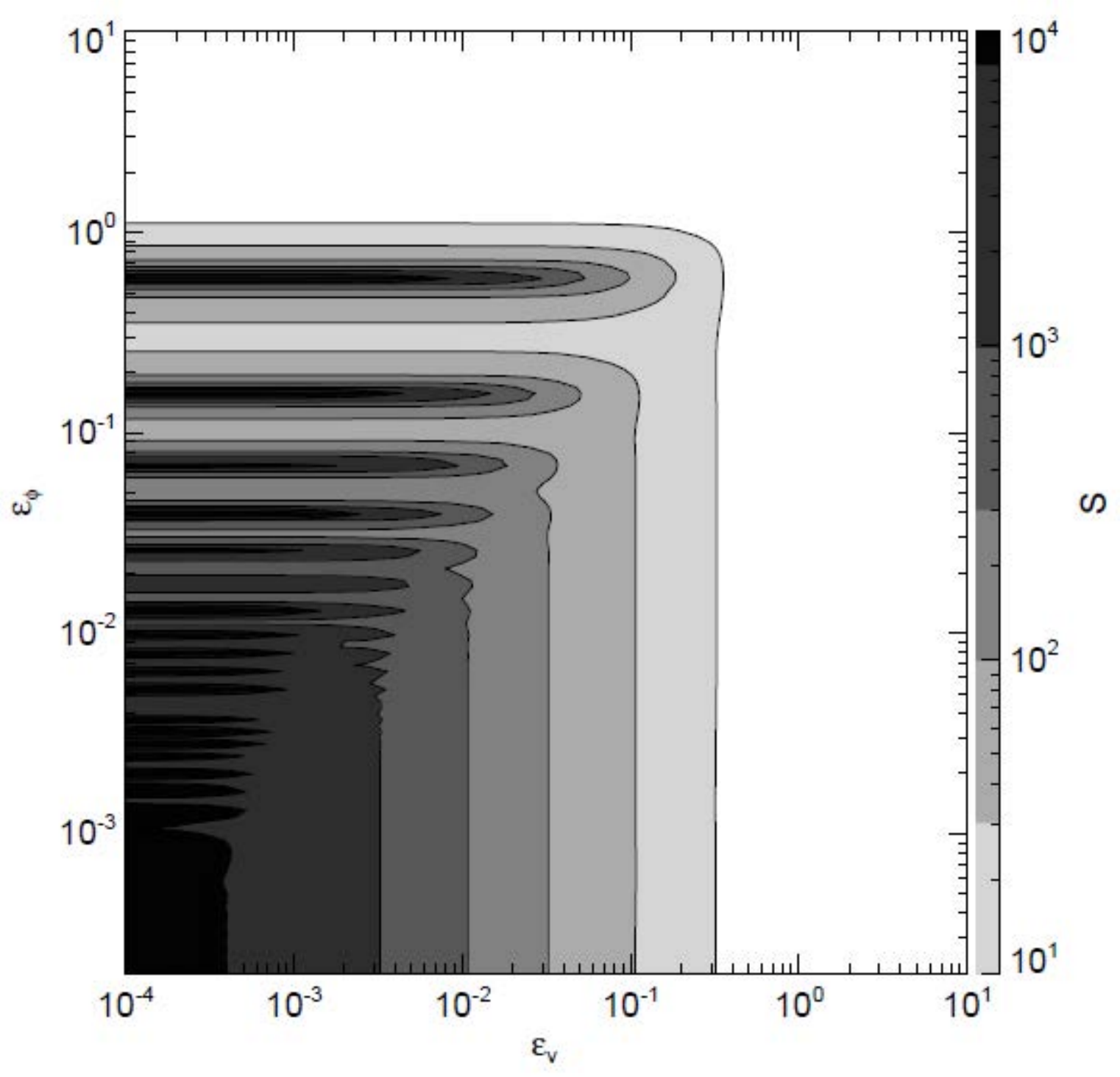}
\caption{Sommerfeld boost in the $\varepsilon_\phi-\varepsilon_\nu$ plane \cite{ArkaniHamed:2008qn}.}
\label{fig:Sommerfeld}
\end{figure}

In Fig.~\ref{fig:CrossSection} the different contributions to the 
total DM annihilation cross section are shown according to Eq.~\eqref{eqn:sigma}.
It can be noted that the Sommerfeld contribution is a function of 
the redshift $z$ but, after a particular value of $\nu$ ($\sim10^{-4}c$), which is very close to the one 
expected for our dark halo, the quantum effect reaches a plateau that is approximately 
100 times the thermal annihilation cross section, \textit{i.e.} $3\times10^{-24}\ cm^3\ s^{-1}$. Here one can notice 
that there should be necessarily a slow velocity limit, given by the finite range of the 
attractive force mediated by $\phi$: once the de Broglie wavelength of the particle 
$(m_\textup{DM}\nu)^{-1}$ exceeds the range of the interaction $m_\phi^{-1}$, the quantum effect 
saturates reaching the value $S_{max}\sim\alpha_\chi m_\textup{DM}/m_\phi$, that is a constant and 
does not depend on $\nu$ \cite{Slatyer:2009yq}. 
Moreover, for specific values of $\varepsilon_\phi$ and $\varepsilon_\nu\equiv\beta/\alpha_\chi$, 
where $\beta$ is the ratio of $\nu$ to the speed of light $c$, resonant threshold states can be 
produced (see Fig.~\ref{fig:Sommerfeld}) which are capable of further 
boosting $\langle\sigma\nu\rangle$ \cite{ArkaniHamed:2008qn}, inducing $S\ge 10^3$.

After cosmological recombination, DM had 
a velocity proportional to the inverse of 
the scale factor of the expanding Universe because of its free stream, justifying very small velocity 
values; on the other hand, the virialized velocities in a today galactic halo might be at least 
three order of magnitude greater.\\
The status of the three cross sections in the previously mentioned regimes can be basically 
summarized as follows:
\begin{enumerate}[1.]
\item \textit{Freeze-out}. During this phase the cross section is usually assumed to be nearly constant, 
without the general 
contribution of other even powers of $\nu$. This is an absolute minimum from which the annihilation 
cross section evolves during the cooling of the Universe. As shown in Fig.~\ref{fig:CrossSection}, 
for ''too relativistic'' freeze-out velocities, the p-wave term would enhance the 
annihilation cross section of one order of magnitude, generating an incompatibility w.r.t. 
the expected relic density at redshift $z=0$. For this regime the relative velocity is:
\be
\nu|_{z \gg 1000} = \nu_\textup{freeze-out} \lesssim 0.3 c
\ee
and the following relation for the cross section holds:
\be
\langle\sigma\nu\rangle^\textup{ann}_\textup{therm} 
\sim 3\times10^{-26}\ cm^3\ s^{-1}\,.
\ee

\item \textit{Recombination}. Here the annihilation cross section has a greater value than the 
freeze-out one, \textit{i.e.} $\langle\sigma\nu\rangle_\textup{rec}>\langle\sigma\nu\rangle_\textup{therm}$. This 
does not imply the relation 
$\langle\sigma\nu\rangle_{rec}>\langle\sigma\nu\rangle_\textup{halo}$, 
because the quantum Sommerfeld enhancement saturates after a certain velocity $\nu_\textup{sat}$, 
which is $\nu_\textup{sat}\lesssim 10^{-3}c$. So, for $\nu<\nu_\textup{sat}$ the quantum Sommerfeld boost $S(z)$ 
does not increase the annihilation cross section further. 
This is the annihilation cross section measured by CMB experiments. In this 
regime the relative velocity of the DM particles is very low:
\be
\nu|_{600<z<1000} = \nu_\textup{rec} \sim 10^{-(8\div6)}c
\ee
and the annihilation cross section is clearly dominated by the Sommerfeld enhancement:
\be
\langle\sigma\nu\rangle^\textup{ann}_\textup{rec} = 
S_\textup{rec}\langle\sigma\nu\rangle_\textup{therm}\,.
\ee

\item \textit{Galactic halo}. As for recombination epoch, the dominant term is the Sommerfeld one. 
In our Galaxy the velocity of the DM particles is supposed to be of the order of 
($100\div300)\ km\ s^{-1}$, which implies $\nu \lesssim 10^{-3}c$. In the Galaxy the velocity regime is:
\be
\nu|_{z=0} = \nu_\textup{halo} \sim 5\times10^{-4}c
\ee
with an annihilating cross section:
\be
\langle\sigma\nu\rangle^\textup{ann}_\textup{halo} 
= S_\textup{halo}\langle\sigma\nu\rangle_\textup{therm}\,.
\ee
\end{enumerate}
If we take into account that the current value of $\nu$ in our Galaxy does not completely saturate the 
Sommerfeld effect, a conservative numerical factor $2$ can be assumed for the ratio $S_{rec}/S_{halo}$
between the enhancement at recombination and the one today (see Fig.~\ref{fig:CrossSection}). The 
relation between the galactic annihilation cross section measured by AMS-02 and the one measured by 
Planck at the recombination can be written as:
\be
\label{eqn:CrossSection}
\langle\sigma\nu\rangle^\textup{ann}_\textup{rec}\approx(1\div2)\langle\sigma\nu\rangle^\textup{ann}_\textup{halo}\,.
\ee

\section{Constraints on dark matter properties from astroparticle physics and cosmology}
\label{sec:three}

The CMB experiments can constrain the DM annihilation cross section from the quantity 
introduced in Refs.~\cite{Chen:2003gz,Padmanabhan:2005es}:
\be
p_\textup{ann}\equiv f_\textup{eff}\frac{\langle\sigma\nu\rangle_\textup{rec}}{m_\textup{DM}}
\ee
as a function of the factor $f_\textup{eff}$ that encodes the efficiency of the energy absorption 
at the recombination. Therfore, another parameter has to be taken into account for the comparison 
between CR and CMB experiments. In principle $f_\textup{eff}=f(z)$, 
but it has been demonstrated that this can be taken as a constant at $z\simeq600$ 
\cite{Galli:2011rz,Finkbeiner:2011dx}, around the recombination epoch; it lies 
in the theoretical range between $0.01$ and $1$, with recombination values generally chosen 
between $0.12$ and $0.6$ as a function of the annihilation channel (see 
Refs.~\cite{ArkaniHamed:2008qn,Galli:2009zc,Finkbeiner:2011dx}). 
So, for small $f_\textup{eff}$ values the annihilation  constraints are relaxed.

\begin{figure}[htp!!]
\centering
\includegraphics[height=6cm]{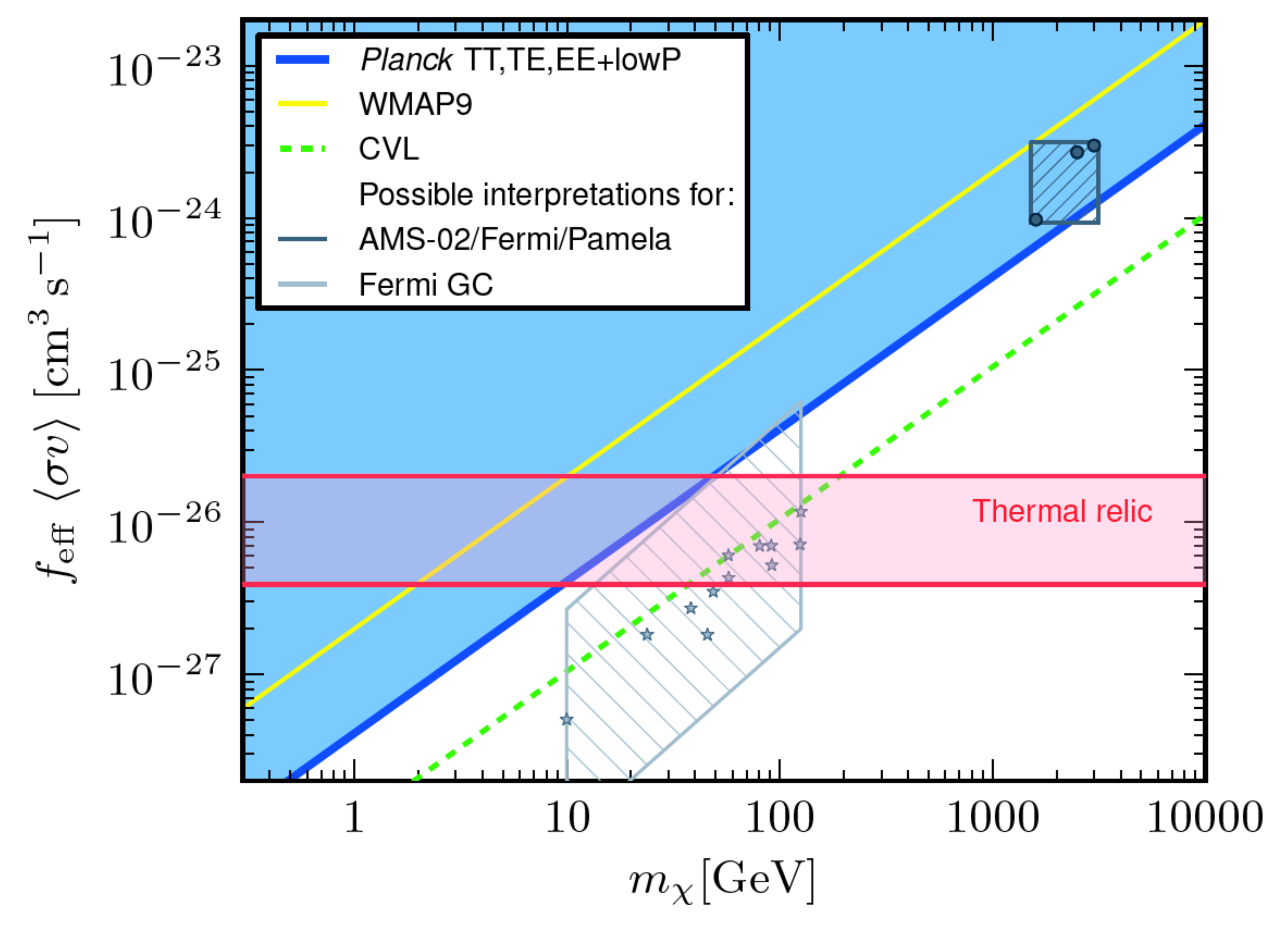}
\caption{Annihilation cross section constraint as a function of DM mass.
(Image credit: ESA and the Planck Collaboration, \cite{Ade:2015xua}.)
The AMS-02/Fermi/PAMELA rectangle was obtained from Ref.~\cite{Cholis:2013psa} 
considering ($m_\textup{DM}=1.6\ TeV$, $\langle\sigma\nu\rangle=6.5\times10^{-24}\ cm^3\ s^{-1}$, 
$f_\textup{eff}=0.15$), ($m_\textup{DM}=2.5\ TeV$, 
$\langle\sigma\nu\rangle=1.5\times10^{-23}\ cm^3\ s^{-1}$, $f_\textup{eff}=0.18$) and 
($m_\textup{DM}=3\ TeV$, $\langle\sigma\nu\rangle=2.3\times10^{-23}\ cm^3\ s^{-1}$, 
$f_\textup{eff}=0.13$) for $2\mathit{e}/2\mathit{\mu}/2\mathit{\pi}$, 
$4\mathit{\mu}$ and  $4\mathit{\pi}$ respectively.\protect\footnotemark}
\label{fig:PlanckAnnihilation}
\end{figure}

In Refs.~\cite{Galli:2009zc,Slatyer:2009yq} have been obtained the following  bounds for the saturated 
annihilation cross section and the Sommerfeld factor from WMAP 5 yr data:
\begin{align}
\label{eqn:5cross}
\langle\sigma\nu\rangle_\textup{sat} &< \frac{3\times10^{-24}\ cm^3\ s^{-1}}{f_\textup{eff}} 
\left(\frac{m_\textup{DM}}{1\ TeV}\right) \\
\label{eqn:5enhancement}
S_\textup{max} &< \frac{120}{f_\textup{eff}}\left(\frac{m_\textup{DM}}{1\ TeV}\right)\,.
\end{align}
For $1\ TeV$ DM candidate and $f_\textup{eff}=0.12$ Eqs.~\eqref{eqn:5cross}-\eqref{eqn:5enhancement} lead to 
$\langle\sigma\nu\rangle_\textup{sat}<2.5\times10^{-23}\ cm^3\ s^{-1}$ and $S_\textup{max}<10^3$.\\
Now, with the latest Planck 2015 data \cite{Ade:2015xua} this constraint can be 
improved of about one order of magnitude (see in Fig.~\ref{fig:PlanckAnnihilation} the comparison between WMAP9 and Planck constraints), 
leading, for the previous case, to $\langle\sigma\nu\rangle_\textup{sat} < 10^{-24}\ cm^3\ s^{-1}$ and $S_\textup{max} < 10^2$.\\
\footnotetext{We thank Silvia Galli for clarifying this point.}
In Fig.~\ref{fig:PlanckAnnihilation} the constraints from Planck 
\cite{Ade:2015xua}, obtained with the full temperature data and the inclusion of low- and high-$l$ 
polarization data, are shown. Here, the allowed region of parameters space 
for charged CR measurements was taken from 
Cholis and Hooper \cite{Cholis:2013psa}, 
under the assumption that the CR positron excess was due to pure DM annihilation.\\
From Fig.~\ref{fig:PlanckAnnihilation} it can be noted that, for a TeV-ish particle, \textit{e.g.} 
with a mass of $3\ TeV$, and a \textit{pessimistic case} with an ideal absorption 
efficiency $f_\textup{eff}=1$, the annihilation cross section at 
recombination from Planck is $\langle\sigma\nu\rangle_\textup{rec} < 10^{-24}\ cm^3\ s^{-1}$,
which implies 
$\langle\sigma\nu\rangle_\textup{halo} < 0.5\times10^{-24}\ cm^3\ s^{-1}$ at most. Instead, in 
an \textit{optimistic case} 
with $\langle\sigma\nu\rangle_\textup{rec}\approx\langle\sigma\nu\rangle_\textup{halo}$ and 
$f_\textup{eff}=0.1$, 
the constraint for a TeV-ish DM becomes $\langle\sigma\nu\rangle<10^{-23}\ cm^3\ s^{-1}$.\\
Several studies suggest values less than $0.4$ for $f_\textup{eff}$ at recombination, that is a rather 
optimistic scenario for indirect search constraints \cite{Slatyer:2015jla,Slatyer:2015kla}.
Consequently, constraints of the order of $10^{-(24\div23)}\ cm^3\ s^{-1}$ for a heavy DM 
candidate can be still in 
agreement with the boosted cross sections which are necessary to 
reproduce CR positrons excess \cite{Masi:2013phd}; as a first step, only resonant Sommerfeld boosts, that induce an 
overall enhancement $>10^3$  w.r.t $\langle\sigma\nu\rangle_\textup{therm}$, can be for certain excluded.

It follows that it is interesting and 
necessary to discuss and update the small rectangle in the right upper 
corner of Fig.~\ref{fig:PlanckAnnihilation}, related to 
AMS-02 2013 \cite{Aguilar:2013qda}, Fermi \cite{Ackermann:2010ij} and  PAMELA \cite{Adriani:2011xv} data, 
in light of new CR results and speculations, to avoid falling into misleading conclusions.\\
In fact, the analysis in \cite{Cholis:2013psa} was performed with only 2013 AMS-02 data 
along with PAMELA, Fermi and AMS-01 ones. 
It must be also noted that Fermi all electrons channel  $\mathit{e^+}+\mathit{e^-}$ is in tension with 
AMS-02 positron fraction, as 
stressed in \cite{Boudaud:2014dta}, and it has not 
been reprocessed and confirmed after the recent pass 8 calibration performed in \cite{TheFermi-LAT:2015hja} 
\footnote{Peter Michelson, Fermi-LAT PI, confirms the non reliability of Fermi 
$\mathit{e^+}+\mathit{e^-}$ data (and consequently of positron fraction and $\mathit{e^+}$ data)
at the AMS-days at CERN on April 2015, see
\href{https://cds.cern.ch/record/2010841}{https://cds.cern.ch/record/2010841} at $18'$.}.\\
Furthermore, putting all these CR data together could produce some inconsistencies and issues: since 
the data used for fits were collected by different instruments, the fits errors are 
hard to estimate.

\begin{figure}[htp]
\centering
\includegraphics[height=6cm]{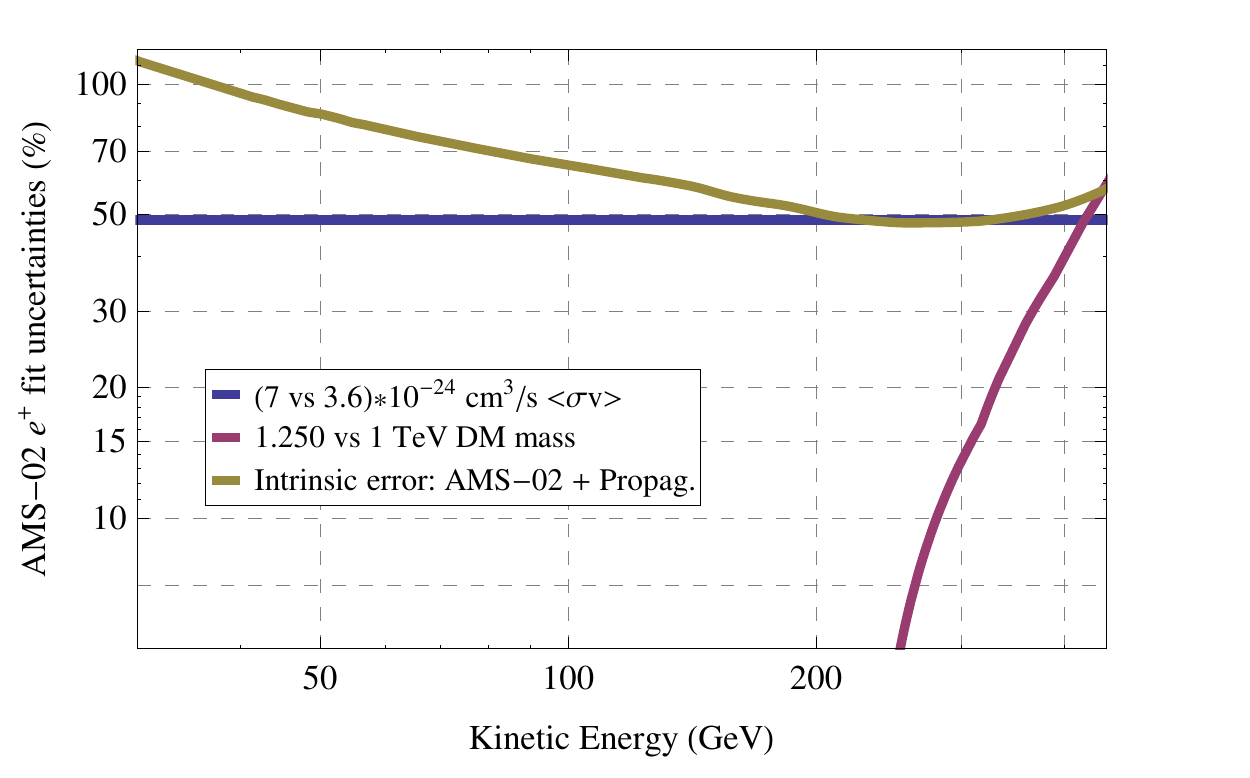}
\caption{Uncertainties comparison associated to cosmic positron spectrum measured by AMS-02. 
We plot the ratio between a reference positron fit with 
$\langle\sigma\nu\rangle=7\times10^{-24}\ cm^3\ s^{-1}$ and a set with about half the 
annihilation cross section (blue curve) and the ratio between a reference positron fit 
with $m_\textup{DM}=1\ TeV$ and one with $m_\textup{DM}=1.25\ TeV$ (purple curve). As long the two curves
lie below the dark yellow line, which represent the combination 
of the two main uncertainties sources, from CR propagation 
(the min-max sets span in \cite{Cirelli:2010xx}) and from AMS-02 errors, the 
best-fit point can be adjusted in the $\langle\sigma\nu\rangle-m_\textup{DM}$ plain.}
\label{fig:PPPC4}
\end{figure}

The most interesting constraints in literature for TeV-ish DM candidates, \textit{i.e.} 
$m_\textup{DM}\ge1\ TeV$, using AMS-02 positron data are the following:
\begin{enumerate}[1.]
\item From the previously discussed Ref.~\cite{Cholis:2013psa} it must be noted that the reduced $\tilde{\chi}^2$ for 
TeV-ish DM fits is much better for Fermi data rather than for AMS-02 2013 ones. The good fits, 
with $\tilde{\chi}^2\sim1$, for AMS-02 data in \cite{Aguilar:2013qda}, for a DM which 
annihilates into a pair of intermediate states $\phi$ are  
$\langle\sigma\nu\rangle=5.8\times10^{-24}\ cm^3\ s^{-1}$ with 
$m_\textup{DM}=1\ TeV$, $\langle\sigma\nu\rangle=6.5\times10^{-24}\ cm^3\ s^{-1}$ with 
$m_\textup{DM}=1.6\ TeV$, $\langle\sigma\nu\rangle=1.5\times10^{-23}\ cm^3\ s^{-1}$ with 
$m_\textup{DM}=2.5\ TeV$ and $\langle\sigma\nu\rangle=2.3\times10^{-23}\ cm^3\ s^{-1}$ with 
$m_\textup{DM}=3\ TeV$ for $4\mu$, $4\pi$ and $2\mathit{e}/2\mathit{\pi}/2\mathit{\mu}$ combinations. The overall result 
is a DM with a mass up to $3\ TeV$ and an annihilation cross section in the range of few 
$10^{-(24\div23)}\ cm^3\ s^{-1}$;

\item In Ref.~\cite{Boudaud:2014dta} some fits to AMS-02 2014 positron fraction data 
with TeV-ish DM are obtained: $\langle\sigma\nu\rangle \approx 4.5\times10^{-23}\ cm^3\ s^{-1}$ 
with $m_\textup{DM}\approx1.76\ TeV$ for $4\tau$ channel, 
$\langle\sigma\nu\rangle \approx (2.5\div7)\times10^{-23}\ cm^3\ s^{-1}$ with 
$m_\textup{DM}\approx(1\div2)\ TeV$, with little deviations as a function of the annihilation 
channels combinations. Some fits are achieved using also Fermi-LAT data, which 
have a great uncertainty w.r.t. AMS-02 2014 data.

\item In Ref.~\cite{Cirelli:2008pk} the privileged regions, using AMS-02 2013 data, are about 
$m_\textup{DM}\approx1\ TeV$ with 
$\langle\sigma\nu\rangle \approx 10^{-23}\ cm^3\ s^{-1}$. They also used Fermi and HESS data;

\item In \cite{Feng:2013zca}, using AMS-02 2013 data, along with PAMELA and Fermi ones, 
they obtain $m_\textup{DM}\approx1\ TeV$ with $\langle\sigma\nu\rangle \approx 6\times10^{-24}\ cm^3\ s^{-1}$;

\item In Ref.~\cite{Liu:2013vha}, using AMS-02 2013, the results for a Sommerfeld boost $S$ from a scalar, 
pseudoscalar or vector particle are approximately the same: $m_\textup{DM}\approx1\ TeV$ with 
$S \approx 1.5\times10^2$ and 
$\langle\sigma\nu\rangle=S\langle\sigma\nu\rangle_\textup{therm} \approx 4.5\times10^{-24}\ cm^3\ s^{-1}$ 
for the $4\mathit{\mu}$ channel, $m_\textup{DM}\approx(1.5\div2)\ TeV$ with 
$S \approx (5\div9)\times10^2$ and 
$\langle\sigma\nu\rangle=S\langle\sigma\nu\rangle_\textup{therm}
\approx(1.5\div2.7)\times10^{-23}\ cm^3\ s^{-1}$ for the $2\mathit{\tau}$ channel, 
$m_\textup{DM}\approx(3\div4)\ TeV$ with 
$S \approx (1\div2)\times10^3$ and 
$\langle\sigma\nu\rangle=S\langle\sigma\nu\rangle_\textup{therm}
\approx(3\div6)\times10^{-23}\ cm^3\ s^{-1}$ for the $4\mathit{\tau}$ channel;

\item In Ref.~\cite{Ibarra:2013zia}, from AMS-02 2013 positron fraction data the upper limit of 
$\langle\sigma\nu\rangle \approx 10^{-(24\div23)}\ cm^3\ s^{-1}$ is achieved, as a function of 
the annihilation channels, associated to a $m_\textup{DM} = 1\ TeV$. They also use AMS-02 positrons 
from ICRC 2013, which were not finalized nor official data;

\item In Ref.~\cite{Lin:2014vja}, the best-fits obtained from AMS-02 2014 data are about 
$m_\textup{DM}=1(4)\ TeV$ with $\langle\sigma\nu\rangle \approx 0.5(7)\times10^{-23}\ cm^3\ s^{-1}$, 
for $\mathit{\mu}$ and $\mathit{\tau}$ channels;

\item In Ref.~\cite{Lopez:2015uma}, the AMS-02 2014 data lead to a best-fit via Sommerfeld boson that 
is of the order of $m_\textup{DM}=(1\div2)\ TeV$ with 
$\langle\sigma\nu\rangle \approx (1\div5)\times10^{-23}\ cm^3\ s^{-1}$, for $4\mathit{\mu}$ and 
$\mathit{\tau}$ channels.
\end{enumerate}
These studies share common properties and weaknesses. First of all the allowed 
and privileged 
leptonic annihilation channels for DM with mass $1\ TeV\le m_\textup{DM}<10\ TeV$ are generally 
the muonic and tauonic ones. Also annihilation into quarks $\mathit{u,\ d,\ b}$ 
\cite{Boudaud:2014dta} and $\pi$ \cite{Cholis:2013psa} are suitable for heavy DM. 
These best-fits are usually computed with DM masses $m_\textup{DM}\gtrsim1\ TeV$, 
however mass values in the 
$(4\div10)\ TeV$ range, the most interesting one for a heavy WIMP scenario 
\cite{Masi:2015fka}, are poorly tested in literature; on the other hand, some studies are performed for very heavy DM 
with $m_\textup{DM}>10\ TeV$, which is disfavored by recent astrophysical observations 
\cite{Masi:2015fka}. In addition, the greater the mass, the greater the uncertainty of the annihilation 
scheme and the greater the degrees of freedom to tune the annihilation chain and fit the data.\\
It must be stressed that all recent observations 
point toward a TeV-ish paradigm and suggest to look above $1\ TeV$ for WIMP DM masses \cite{Masi:2015fka}; 
at the same time too high DM masses $\ge10\ TeV$ could introduce some conflicts and could be not 
too suitable to fit the AMS-02 2014 positron fraction and its supposed flattening, due to 
the achievement of a maximum of the positron production. It is more advisable to avoid 
the $\mathit{t,\ h,\ W^{\pm}}$ annihilation scenarios associated to very high masses and too high 
annihilation cross sections \cite{Boudaud:2014dta}.\\
The previous fits have one order of magnitude of span in the annihilation cross section; the 
ensemble of the fits prescriptions could be approximately described as a rectangle in the 
$\langle\sigma\nu\rangle-m_\textup{DM}$ plane: $(5\times10^{-(24\div23)}\ cm^3\ s^{-1})\times(1\div4\ TeV)$. 
This must be translated in the $f_\textup{eff}\langle\sigma\nu\rangle-m_\textup{DM}$ plane: if we take 
$0.1\le f_\textup{eff}\le1$, the most general permitted window becomes 
$(5\times10^{-(25\div23)}\ cm^3\ s^{-1})\times(1\div4\ TeV)$.\\
But such a rectangle is very loose and it is based on standard assumptions which 
can be easily extended. In fact the constraints from AMS-02 positron data greatly relax introducing a 
Dark Disk (DD) in addition to a dark halo: this  ensures dynamical enhancements of the 
DM annihilation cross section within the high density DD, $\rho_\textup{DDDM}\gg0.4\ GeV\ cm^{-3}$, 
without the need of $\langle\sigma\nu\rangle>10^{-24}\ cm^3\ s^{-1}$, \cite{Fan:2013yva}. 
For indirect detection, the DD scenario could easily accommodate a large boost factor from local 
density enhancement in the range $10\div1000$, depending on the disk height.\\
In addition, the constraints become less stringent if one consider both DM and 
pulsars as sources of primary positrons \cite{Fend:2015uta}. Pulsars and DD hypothesis can only relax 
the DM bounds, enlarging the allowed region in the $f_\textup{eff}\langle\sigma\nu\rangle-m_\textup{DM}$ 
plane.\\
For what concerns the uncertainties which afflict CR propagation physics, we still do not have a complete 
and well-posed understanding of the CR lepton problematic: non-standard propagation models may be introduced to account for part or all of the 
positrons measured in space \cite{Mertsch:2014poa}). Hence the constraints obtained from the positron sector (positron spectrum and positron fraction) 
strongly depend on the underlying CR propagation model and certainly on the AMS-02 measurements errors. 
Thanks to  AMS-02 unprecedented precision it will be soon possible to fix an almost univocal 
scheme of propagation of cosmic rays in our galaxy, allowing to correctly compute the effective 
background for DM indirect searches: in fact, tiny variations of the most significant parameters, 
such as the diffusive halo thickness, the diffusion coefficient and exponent and the 
electrons spectral indices, may lead to misleading interpretations of AMS-02 data, pointing to an 
incorrect scenario. In Fig.~\ref{fig:PPPC4} a qualitative illustration of propagation and 
measurement uncertainties, performed with PPPC4DMID 
\cite{Cirelli:2010xx}, show how unstable these fits are: 
a best-fit point in the $\langle\sigma\nu\rangle-m_\textup{DM}$ plain carries up to 50\% 
uncertainty in the choice of the annihilation cross section and DM mass values.

From Fig.~\ref{fig:PPPC4} one argues that DM fits to AMS-02 positrons and positron 
fraction with standard propagation models, up to $(400\div450)\ GeV$, suffer large uncertainties, of the order of $(25\div50)\ \%$, 
if taken individually. 
This is the degree of uncertainty from the most precise space experiment which measures CR fluxes: when a fit on PAMELA, Fermi 
(or AMS-01, HEAT, HESS) data is performed, an uncertainty at least three times the AMS-02 ones 
should be addressed: it could imply almost one order of magnitude in the annihilation cross 
section and more than $1\ TeV$ for TeV-ish DM masses.\\
Finally, when exploring the DM $\langle\sigma\nu\rangle-m_\textup{DM}$ space, other underlying parameters are 
fixed, such as the dark matter halo shape and the next-to-leading order (NLO) corrections to 
primary positrons from DM annihilation. The choice of the DM radial profile is not too significant, 
whereas full calculations of the NLO and NNLO electroweak (EW) corrections may modify the primary 
DM fluxes up to one order of magnitude \cite{Ciafaloni:2012gs,DeSimone:2012np,Hryczuk:2011vi}, producing more 
light final states than expected in LO calculations and allowing lower values of the annihilation 
cross section: they are relevant for spectra predictions especially when $m_\textup{DM}$ is much larger 
than the EW scale and EW bremsstrahlung is permitted.

\section{Results and Conclusions}
\label{sec:conclusion}

\begin{figure}[htp]
\centering
\includegraphics[width=9cm]{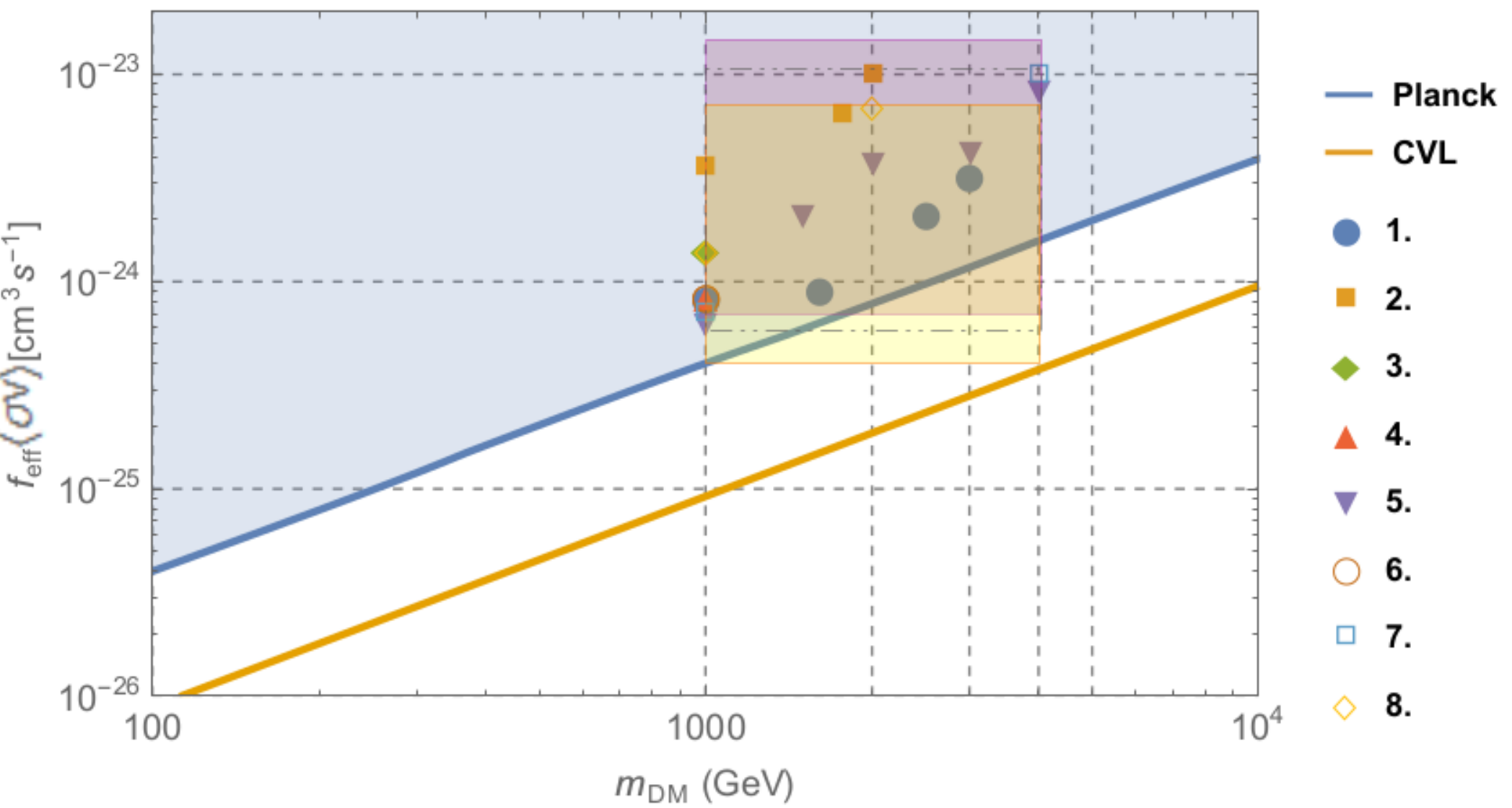}
\caption{Allowed parameters space for indirect search on $f_\textup{eff}\langle\sigma\nu\rangle-m_\textup{DM}$ 
plane from AMS-02/Fermi/PAMELA (dot-dashed rectangle), along with Planck 2015 bound (blue line). 
Only a Cosmic Variance Limited (CVL) experiment (black line) could completely falsify the CR 
indirect search constraints for TeV-ish candidates. The points corresponding to the best-fits are the ones presented 
in Section~\ref{sec:three} for $f_\textup{eff}=0.15$. The rectangles 
for $f_\textup{eff}=0.1$ (yellow) and $f_\textup{eff}=0.2$ (purple) are also drawn.}
\label{fig:PlanckAnnihilationRev}
\end{figure}

We consider the collection of fits discussed in Section~\ref{sec:three}: applying the constraint  
$0.1< f_\textup{eff}<0.4$, according to the up-to-date calculations derived in 
\cite{Slatyer:2015jla,Slatyer:2015kla}, where $0.4$ is preferred only for the electron channels for non TeV-ish DM candidates, 
an overall CR allowed region is obtained 
(Fig.~\ref{fig:PlanckAnnihilationRev}), which is something like 
$(4\times10^{-25}\div1.5\times10^{-23}\ cm^3\ s^{-1})\times(1\div4\ TeV)$. 
The qualitative fraction of this region permitted by CMB observations is about 
9\%, 4\% and 2\% of the total area obtained combining cosmic rays fits, assuming $f_\textup{eff}=0.1,\ 0.15,\ 0.2$ respectively.\\
Our analysis is compatible with the one presented in \cite{Ade:2015xua}; in 
\cite{Ade:2015xua} the region allowed by Planck is $\sim2$\%.
\\Planck 2015 CMB data appear not to completely exclude annihilating DM as primary source for AMS-02 positron excess, regardless the choice of the recombination efficiency value. 
From Fig.~\ref{fig:PlanckAnnihilationRev} it emerges that future improvement on CMB 
measurements \cite{Madhavacheril:2013cna} could hardly falsify the dark matter interpretation of cosmic-rays 
antiparticles, for what concerns TeV-ish dark matter candidates. In addition, 
fundamental uncertainties from CR propagation physics and from DD hypothesis, pulsars contributions and alternative CR 
propagation models, not reported in Fig.~\ref{fig:PlanckAnnihilationRev}, could almost arbitrarily enlarges the window in the 
$f_\textup{eff}\langle\sigma\nu\rangle-m_\textup{DM}$ plane, up to $10\ TeV$ particles and down 
to nearly thermal cross sections. Incoming CR high statistics measurements of leptons and nuclei 
from AMS-02 will allow us to deepen the examination of this important issue, narrowing CR uncertainties 
and DM properties. Besides that, the information from the antiproton channel is mandatory to put more consistent and coherent constraints on the annihilation capability of the DM candidate, because CR antiprotons have less important backgrounds \cite{Masi:2013phd}, so granting more precise estimations. A cross check based on a hadron-lepton channels comparison will improve our understanding of DM annihilation and probably enlarge the analysis up to the $10\ TeV$ scale.

\section*{Acknoledgements}
We wish to thank Fabio Finelli and Daniela Paoletti for useful comments and suggestions on the draft. 
MB acknowledge support by the "ASI/INAF Agreement 2014-024-R.0 for the Planck LFI Activity of Phase E2".

\section*{References}

\end{document}